\documentclass[
reprint,
superscriptaddress,
 amsmath,amssymb,
 aps,
prb,
longbibliography,
]{revtex4-2}

\usepackage{graphicx}
\usepackage{dcolumn}
\usepackage{bm}
\usepackage[colorlinks,bookmarks=false,citecolor=blue,linkcolor=green,urlcolor=blue]{hyperref}
\usepackage{color}
\usepackage{xcolor}

\begin{document}

\title{Detecting delocalization-localization transitions from full density distributions}

\author{Miroslav Hopjan}
\affiliation{Institut f\"ur  Theoretische Physik, Georg-August-Universit\"at G\"ottingen, Friedrich-Hund-Platz 1, 37077 G\"ottingen, Germany}

\author{Giuliano Orso}
\affiliation{Université de Paris, Laboratoire Matériaux et Phénomènes Quantiques, CNRS, F-75013, Paris, France}
\author{Fabian Heidrich-Meisner}
\affiliation{Institut f\"ur  Theoretische Physik, Georg-August-Universit\"at G\"ottingen, Friedrich-Hund-Platz 1, 37077 G\"ottingen, Germany}

\begin{abstract}
Characterizing the delocalization transition in closed quantum systems with a many-body localized phase is a key open question in the field of 
nonequilibrium physics.
We exploit that localization of particles as realized in Anderson and standard many-body localization (MBL)
implies Fock-space localization in single-particle basis sets characterized by a real-space index. 
Using a recently introduced quantitative measure for  Fock-space localization computed from 
the density distributions, the occupation distance, we systematically study its scaling behavior across delocalozation transitions and identify critical 
points from scaling collapses of numerical data. Excellent agreement with literature results  is found for the critical disorder strengths of noninteracting fermions, such as  the one-dimensional Aubry-Andr\' e and the three-dimensional Anderson model.
We  observe a distinctively different scaling behavior in the case of  interacting fermions with random disorder
consistent with  a Kosterlitz-Thouless transition. 
Finally, we use our measure  to extract the transition point as a function of filling for interacting fermions.
\end{abstract}

\maketitle

\section{Introduction}
Anderson localization \cite{Anderson65} 
 generalizes to disordered systems of interacting particles \cite{Gornyi05,Basko06} leading to 
 many-body localization (MBL). MBL  conceptually fits into 
the framework of thermalization in closed quantum systems \cite{Deutsch91,Srednicki94,Rigol2008,dAlessio16} 
as a generic exception from eigenstate thermalization \cite{Altman15,Nandkishore15,Altman18,Alet18,Abanin19}. The MBL transition is visible in properties of many-body eigenstates
 at a finite energy density \cite{Nandkishore15}, such as area-law entanglement \cite{Bauer13,Kjall14,Friesdorf15}
 and in properties of time-evolved states in global quenches, such as slow logarithmic entanglement entropy growth
 \cite{Znidaric08,Bardarson12,Serbyn13b} or persistent density inhomogeneities \cite{Schreiber15,Choi16}. 
Experiments with ultracold atoms \cite{Kondov15,Schreiber15,Choi16,Luschen17,Bordia17,Kohlert19,Rubio-Abadal19,Lukin19,Rispoli19,Leonard20}, 
solid-state spin systems \cite{Alvarez15,Wei18}, trapped ions \cite{Smith16}, and
superconducting qubits \cite{Chen17,Roushan17,Chiaro19,Guo20,Gong20} 
emulated various lattice models with disorder and probed their localization properties
\cite{Kondov15,Schreiber15,Choi16,Luschen17,Bordia17,Kohlert19,Rubio-Abadal19,Lukin19,Rispoli19,Leonard20,Alvarez15,Wei18,Smith16,Chen17,Roushan17,Chiaro19,Guo20}.
Some of the experiments measure  eigenstate properties, e.g., the level statistics \cite{Smith16,Roushan17},
while most of the efforts address dynamical aspects including 
the imbalance \cite{Schreiber15,Choi16,Luschen17,Bordia17,Kohlert19,Rubio-Abadal19,Guo20}, 
the time-dependent entanglement entropy \cite{Smith16,Chen17,Lukin19,Chiaro19}, or
$n$-point correlators \cite{Wei18,Rispoli19,Leonard20}. 

There is, however, an ongoing debate on the nature of the ergodic-to-MBL transition and even the very existence of MBL in the field's standard model, namely  interacting spinless fermions in one dimension (1D), has been challenged \cite{Suntajs19,Sierant19a,Abanin19a,Panda20,Suntajs20,Sels20,LeBlond20,Sirker20,BarLev20,Sirker20a,Sirker21}. 
The current proposals for the transition in the thermodynamical limit are:  (i) a continuous transition with a power-law scaling of correlations  \cite{Luitz14,Khemani17}, (ii) a transition involving Griffiths regions \cite{Agarwal15,Lenarcic20}, (iii)  Kosterlitz-Thouless (KT) type scaling \cite{Goremykina19,Dumitrescu19,Morningstar19,Herviou19,Suntajs20,Laflorencie20,Morningstar20,DeTomasi20}
as a special case of (ii) \cite{Lenarcic20}, or  (iv) absence of a true MBL phase in the thermodynamic limit \cite{Suntajs19,Suntajs20,Sels20,LeBlond20,Sirker20,Sirker20a,Sirker21}.
While numerical methods play a key role, they are limited  with regard to the accessible system sizes \cite{Santos04,Oganesyan07,Znidaric08,Pal10,Bardarson12,Kjall14,Luitz14,Friesdorf15,BarLev15,Bera15,Lim16,Khemani16,Khemani17,Alet18,Mace20,Mace20,Laflorencie20,Suntajs19,Sierant19a,Abanin19a,Panda20,Suntajs20,Sels20,LeBlond20,Sirker20,BarLev20,Sirker20a,Sirker21,Hopjan20}. 
Therefore, there is a clear need to identify observables for the characterization of localization-delocalization transitions that can be measured in state-of-the-art 
and future experiments with quantum simulators and are easy to compute numerically as well.

Motivated by these open questions, in this work, we establish the recently introduced occupation distance \cite{Hopjan20} as a
useful quantitative measure for characterizing delocalization-localization transitions. 
As our main results, we first show that the occupation distance   detects
 localization-delocalization transitions in noninteracting {Hamiltonians}, including  the 1D Aubry-Andr\' e  model (AAM) and the three-dimensional (3D) Anderson model (3AM), finding excellent agreement with literature results for the critical disorder strengths. Second, we report evidence for   a KT scaling
for interacting spinless fermions in 1D. Third, as an application,  we study the  dependence of the critical disorder strength on filling in the interacting model.

Given a single-particle basis set $|\phi_\alpha\rangle$ and corresponding 
creation and annihilation operators $c^{\dagger}_\alpha, c_\alpha$, the occupation distance is: 
\begin{equation}
\label{eq:dist}
\delta n_{\alpha}=| n_{\alpha}-[n_{\alpha}] |,
\end{equation}
where $[n_{\alpha}]$ are the closest integer to the occupation $n_{\alpha}=\langle\Psi |c^\dagger_\alpha c_\alpha |\Psi \rangle$ in a given
many-body state $|\Psi\rangle$. In Anderson and MBL insulators, there is Fock-space localization \cite{Basko06,Roy19,Logan19,Bera15} in the basis of quasiparticles 
(l-bits) \cite{Serbyn13b,Huse14,Imbrie16a,Imbrie16b}, the eigenbasis 
of one-particle density-matrices \cite{Bera15,Bera17,Lezama17,Sheng-Hsuan18,Buijsman18,Villalonga19,Chen20,Hopjan20}, the Anderson eigenstates \cite{Bera17}, and in the basis of physical densities $n_{i}=\langle\Psi |c^\dagger_i c_i|\Psi \rangle$ \cite{Bera17,Hopjan20}. 
Here, we concentrate on the latter, since these objects are the easiest to obtain numerically and experimentally. 

\begin{figure}[t!]
\begin{center}
\includegraphics[width=9cm,angle=0]{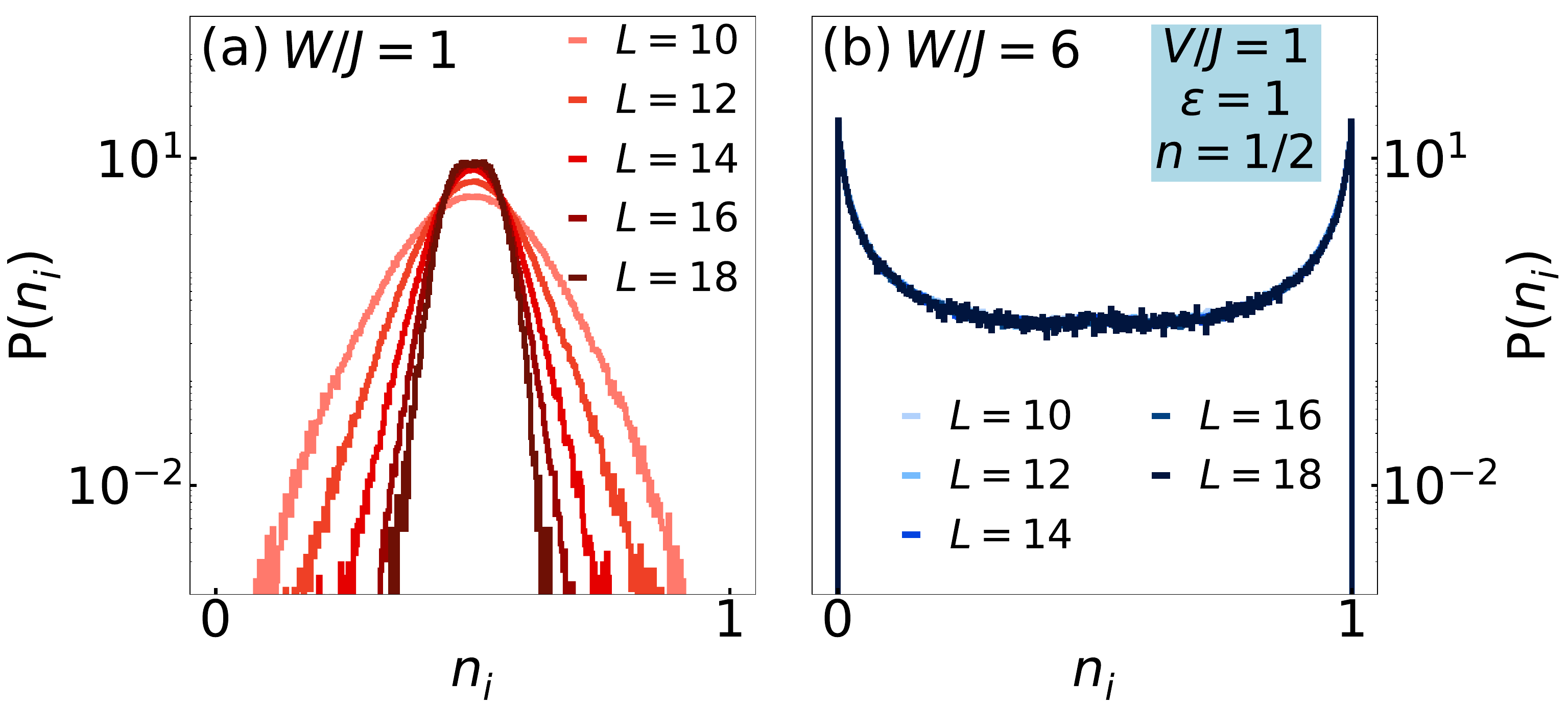}
\caption{Interacting fermions, Eq.~(\ref{eq:ham}), with $V/J=1$, $\epsilon=1$, $n=1/2$: Full distributions 
 of local occupations $n_i$ in (a)  the delocalized  ($W/J=1$) and (b) the localized phase ($W/J=6$) for 
 $L=10, 12, 14, 16, 18$.
 In (b), the distributions overlap.}
\label{histogram}
\end{center}
\vspace{-0.8cm}
\end{figure}

To illustrate the concept of our approach, 
in Fig.$~$\ref{histogram}, we show a typical distribution of $n_{i}$ sampled over eigenstates and disorder realizations for a half-filled chain of interacting spinless fermions \cite{Hopjan20,Luitz2016a,Lim16}, whose model is defined in Eq.~(\ref{eq:ham}) below.
The distribution has a Gaussian-like shape in the delocalized regime [see Fig.$~$\ref{histogram}(a)] and its width shrinks as the system size increases.
 By contrast, in the localized regime [see Fig.$~$\ref{histogram}(b)], the distribution has a bimodal structure and is independent of  system size,
 which is typical for the localized phase.

\begin{figure}[t!]
\begin{center}
\includegraphics[width=9cm,angle=0] {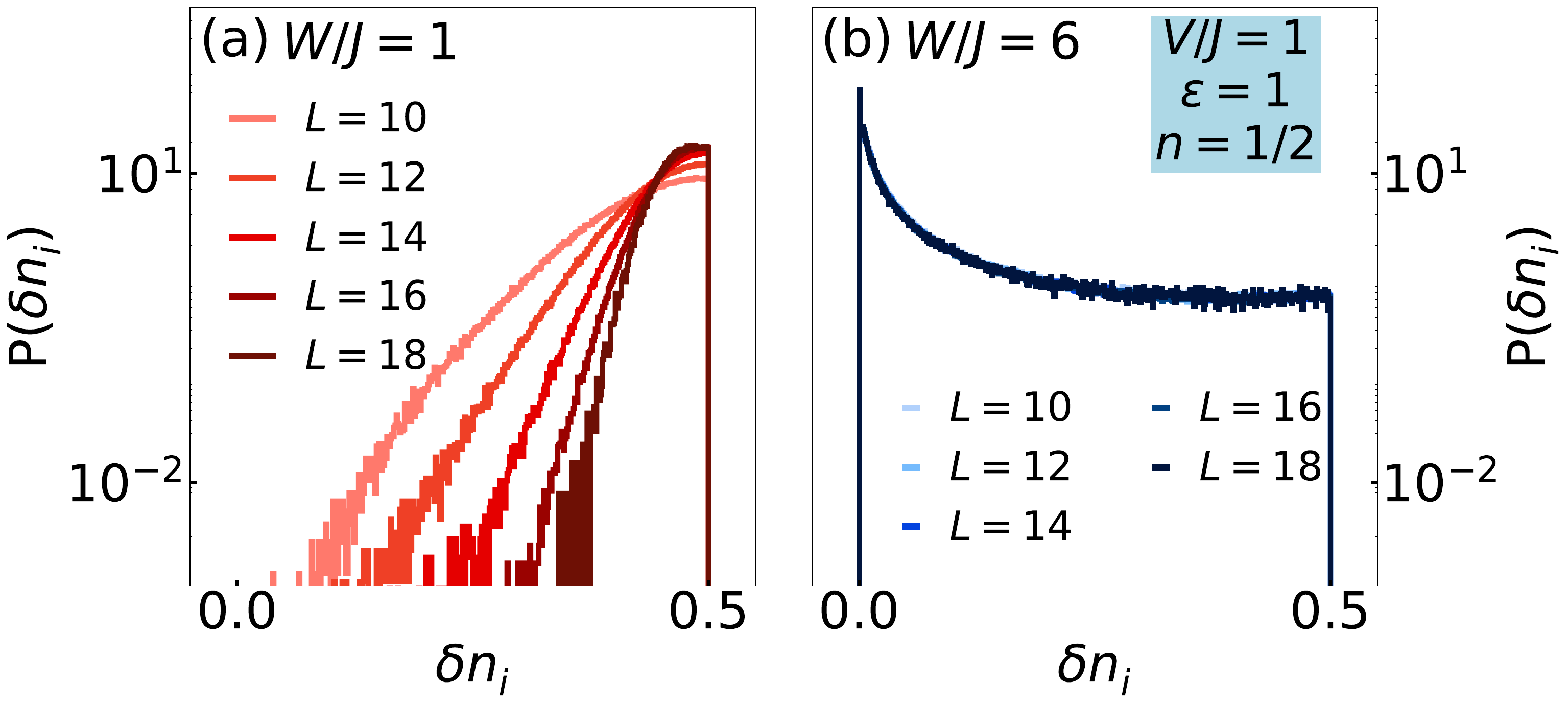}
 \caption{Interacting fermions, Eq.~(\ref{eq:ham}), with $V/J=1$, $\epsilon=1$, $n=1/2$: Full distributions 
of the occupation distancies $\delta n_i$ (a) in the delocalized phase ($W/J=1$) and (b) in the localized phase ($W/J=6$) for 
 $L=10, 12, 14, 16, 18$ are shown.
 For the localized phase in (b), the distributions overlap.}
\label{histogram_sm}
\end{center}
\vspace{-0.8cm}
\end{figure}

In Fig.$~$\ref{histogram_sm}, we show a typical distributions of $\delta n_{i}$  corresponding to the distrubutions of $n_i$ in Fig.$~$\ref{histogram}.
The distribution has a maximum at $\delta n_{i}=n=0.5$ in the delocalized regime [see Fig.$~$\ref{histogram_sm}(a)] which gets larger as the system size increases.
 By contrast, in the localized regime [see Fig.$~$\ref{histogram_sm}(b)], the distribution has a sharp maximum at $\delta n_{i}=0$ and its shape is almost 
 independent of $L$.
{Hence}, in the delocalized regime, the disorder-averaged occupation distance $\overline{\delta n_{i}}$  approaches the average particle filling $n=N/L$  for 
$N,L\rightarrow\infty$ ($N$ being the particle number and $L$ the number of sites), while in the localized regime, it  saturates to a lower value \cite{Hopjan20}.
In the following, we determine the position of the critical point by studying the scaling properties of $n-\overline{\delta n_i}$  on the delocalized  side of   delocalization-localization transitions.

\section{Models}
To establish the validity of our approach, we first  apply it to non-interacting systems given by 
\begin{equation}
\label{eq:ham3DA}
H= -{J}\sum_{\langle ij\rangle} (c_{i}^{\dagger}c_{j}^{}+h.c)+\sum_{i}^{}\epsilon_{i}n_{i}^{}
\end{equation}
that 
exhibit a localization-delocalization transition ($J$ is the hopping matrix element). For the AAM, the
external potential in Eq.~(\ref{eq:ham3DA}) is quasi-periodic, $\epsilon_{i}=W\cos(2\pi q i+\phi)$, where $\phi$ is a global phase and $W$ is
the amplitude of the potential which is incommensurate for an irrational wave number $q$. A standard choice for $q$ is the {inverse} golden ratio
 $q=\frac{\sqrt{5}-1}{2}$. The AAM
has an inherent self-duality at $W_c/J = 2$ giving rise to a sharp metal-insulator transition \cite{Aubry80,Suslov82,Kohmoto83,Chao86,Kohmoto87,Siebesma1987,Hiramoto89,Hiramoto92, Macia2014,Li16,wu2021}, observed
experimentally using cold atoms  \cite{Roati08,Luschen18} and photonic lattices \cite{Lahini:PRL2009}.
In the 3AM, fermions hop on a 3D lattice with
uncorrelated random on-site energies $\epsilon_i \in [-W/2,W/2]$. 
Numerical studies of transport properties \cite{Kramer93,Kramer90,MacKinnon83,MacKinnon81} based on the transfer-matrix technique have shown that, at half filling,
 the system remains insulating for $W >W_{c}  \approx 16.54 J$ \cite{Ohtsuki18} and below $W_{c}$, it
is diffusive \cite{Ohtsuki97,Zhao20}. At the transition, the 3AM exhibits subdiffusion \cite{Ohtsuki97} and
multifractal single-particle wave functions \cite{Rodriguez09,Rodriguez10}. 

To investigate the MBL transition, we consider spinless fermions
with a nearest-neighbor interaction described by the Hamiltonian
\begin{multline}
\label{eq:ham}
H= \sum_{i=1}^L
\Big[-\frac{J}{2}(c_{i}^{\dagger}c_{i+1}^{}+h.c)+\epsilon_{i}(n_{i}^{}-\frac{1}{2})\\ 
+ V (n_{i}^{}-\frac{1}{2})(n_{i+1}^{}-\frac{1}{2})
\Big].
\end{multline}
where $c_{i}^{(\dagger)}$ is a fermionic creation/annihilation operator, $n_{i}=c_{i}^{\dagger}c_{i}$ is the  density at site $i$, $J/2$ is the hopping matrix element, $V$ is 
the strength of the nearest-neighbor 
interactions, and $\epsilon_{i}$ is a random potential drawn from a
uniform box distribution $[-W,W]$ (we use a different convention for $W$ compared to the 3AM to be consistent with the MBL literature).
Using a Jordan-Wigner transformation, 
Eq.~\eqref{eq:ham} maps onto
a spin-$1/2$ XXZ chain with random local magnetic field (note the factor $\frac{J}{2}$ in front of the hopping term), a standard system for studies of MBL~\cite{Oganesyan07,Pal10,Luitz14,Abanin19}. 

 \begin{figure}[t]
\begin{center}
\includegraphics[width=8cm,angle=0]{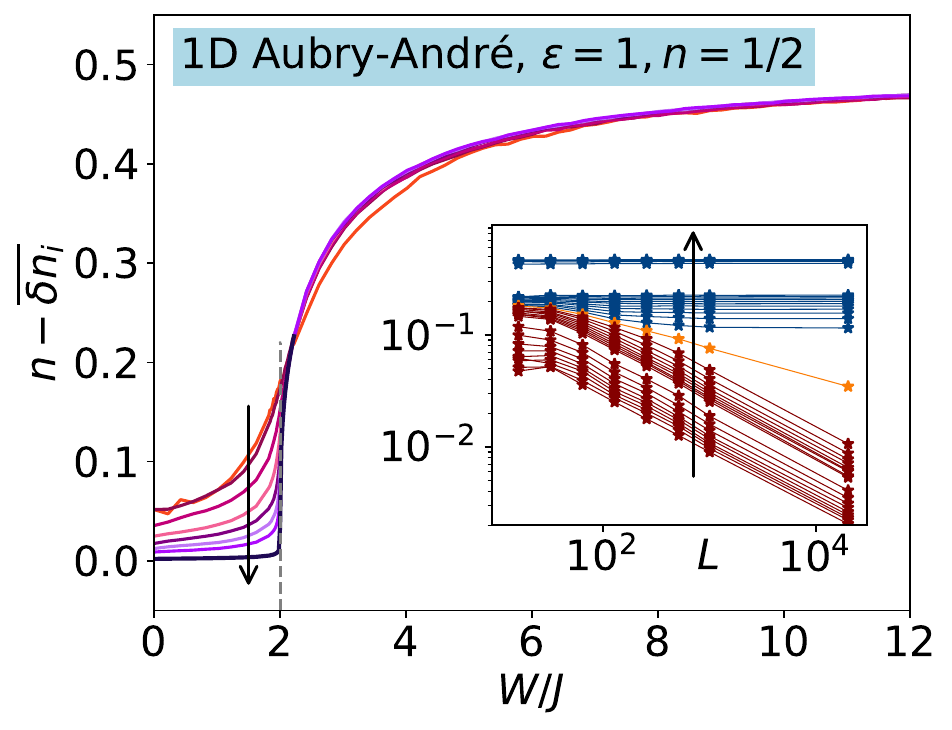}
\caption{AAM: $n-\overline{\delta n_i}$ as a function of $W/J$ for $L=16, 32, 64, 128, 256, 512, 1000, 20000$.
The   arrow   specifies   increasing  $L$ and the dashed line denotes the transition point \cite{Aubry80}. Inset:
Log-log plot of the data from the main panel as a function of $L$ deep
 in the delocalized regime ($W/J= 0.02 -1.62$, in steps of $0.2$, red),  close to the transition  ($W/J= 1.82 -2.22$, in steps of $0.02$, yellow being the transition point),
   and deep in the localized phase  ($W/J= 6.02-12.02$, in steps of $2$, blue). The   arrow   indicates   increasing disorder strength. }
\label{difference_raw_1DAA}
\end{center}
\vspace{-0.8cm}
\end{figure}

We define the target energy density via $\epsilon =\frac{2(E - E_{\rm min})}{E_{\rm max} - E_{\rm min}}$,
where $E$ is the many-body energy of a particular eigenstate and $E_{\rm max}$ and $E_{\rm min}$ are the maximum and minimum
energy for each disorder realization, respectively. Hence $\epsilon=1$ corresponds to the middle of the many-body spectrum.

\section{Results for non-interacting models}
We solve the single-particle problem $H \phi_\alpha= \epsilon_\alpha \phi_\alpha$ with periodic boundary conditions using full exact diagonalization for $L\leq 20000$. 
To obtain those single-particle states $\epsilon_\alpha$ contributing to $\epsilon=1$ many-body states, we use a Monte-Carlo generation to sample the statistics of occupied single-particle orbitals. 
To this end we generate two array $O$ and $U$, containing the indices of  occupied and unoccupied single-particle states, 
respectively (for half filling both arrays have length  $L/2$). 
The initialization of these two arrays is random. Once we have the two arrays, we check if 
the many-body energy $E=\sum_{\alpha \in O} \epsilon_\alpha$ belongs to the target energy window. If not, then we generate a new many-body state by
taking a random integer from each arrays, $r_O$ and $r_U$,  and we exchange the corresponding elements $O_{r_O} \leftrightarrow{} U_{r_U}$.
 We continue until we find a many-body state with
$E$ which belongs to the target energy window. For each disorder realization out of (up to) $10^4$ samples, we take one such many-body eigenstate with $\epsilon \in (0.9995,1.0005)$.

In the main panel of Fig.$~$\ref{difference_raw_1DAA}, we plot $n-\overline{\delta n_i}$  as a function of disorder strength for
different system sizes. 
As expected, in the delocalized phase $(W/J<2)$, this quantity quickly decays
to zero for $L\to \infty$,  while in the localized phase $(W/J>2)$, it saturates to a finite value. This saturation is reached for system sizes  larger
 than the localization length. For the largest system sizes, we observe  saturation  in almost the whole localized phase except for  a
tiny window around the transition, where the localization length diverges. 

 \begin{figure}[t!]
\begin{center}
\includegraphics[width=8cm,angle=0]{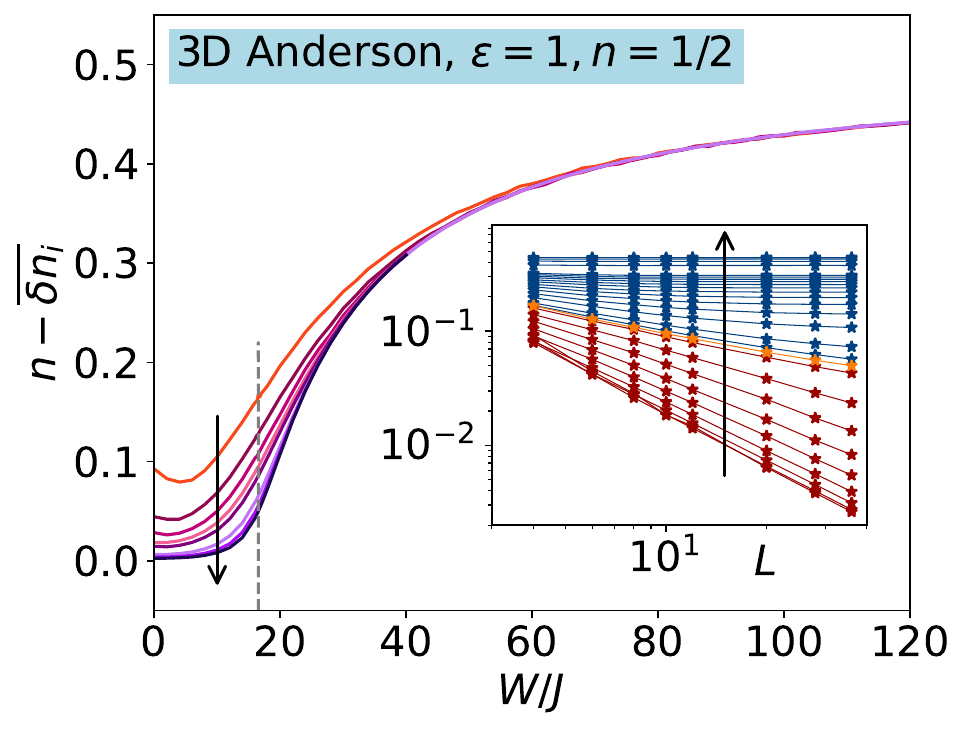}
\caption{3AM, $n-\overline{\delta n_i}$ as a function of $W/J$ for $L^3=4^3, 6^3,8^3, 10^3, 12^3, 20^3, 28^3, 36^3$.
The   arrow   specifies   increasing  system size and the dashed line denotes the transition point estimated from  other
measures $W_c/J\approx 16.54$ \cite{Ohtsuki18}. Inset: Log-log plot of the data from the main panel as a 
function of $L$ in the delocalized regime ($W/J=0.02-16.02 $ in steps of $2$, red), estimated transition point \cite{Ohtsuki18} ($W_c/J\approx 16.54$, yellow) 
and in the localized phase ($W/J= 17.02, 18.02-40.02$  in steps of $2$, blue). The   arrow   specifies   increasing disorder strength.} 
\label{difference_raw_3DA}
\end{center}
\vspace{-0.8cm}
\end{figure}

To get further insights, in the inset of Fig.$~$\ref{difference_raw_1DAA}, we plot $n-\overline{\delta n_i}$  
as a function of $L$ in log-log scale  for different disorder strength $W/J$. 
For $L>200$, the decay is governed by a power law $L^{-\alpha}$
with $\alpha\approx 0.5$ for all values of $W/J<2$. 
Remarkably, exactly at the transition point, the decay is governed by a different exponent,  $\alpha_c\approx 0.25$, revealing
 the different nature of the single-particle wave-functions
 at the critical point \cite{Kohmoto83,Chao86,Kohmoto87,Siebesma1987,Hiramoto89,Hiramoto92,Macia2014,Li16,wu2021}. 
In the localized regime, the saturation is reached already for $L\approx 1000$ even for  values of $W/J$
close to the transition point. To summarize, for the AAM, there is a remarkable
difference of the behavior of the occupation distance comparing the  localized to the delocalized phases.
The power-law decay of $n-\overline{\delta n_i}$ in the metallic phase and its saturation to a constant in the localized 
regime is clearly reminiscent of the behavior of the inverse participation ratio in the single-particle problem.

The corresponding analogue of Fig.$~$\ref{difference_raw_1DAA} for the 3AM is displayed  in Fig.$~$\ref{difference_raw_3DA}.
Here, $n-\overline{\delta n_i}$ clearly decays to zero for $L\to \infty$ in the delocalized phase $(W/J<16)$. 
Oppositely, in the localized regime, there is a clear onset of saturation for $L\approx 20-36$ for 
 $W/J>17$. The trends are less clear for $W_c/J\approx 16-17$, i.e., close to the estimated transition $W_c/J\approx 16.54$ \cite{Ohtsuki18}.
We observe that for all $L$ values considered here, the decay in the delocalized phase is 
governed by a power law $L^{-\alpha}$, but the exponent $\alpha$ now depends on the disorder strength, most probably due to the limited system sizes available. 
At the estimated transition point \cite{Ohtsuki18}, we find that the decay is governed by the exponent $\alpha_c \approx 0.75(5)$.\\

\begin{figure}[t!]
\begin{center}
\includegraphics[width=7.5cm,angle=0]{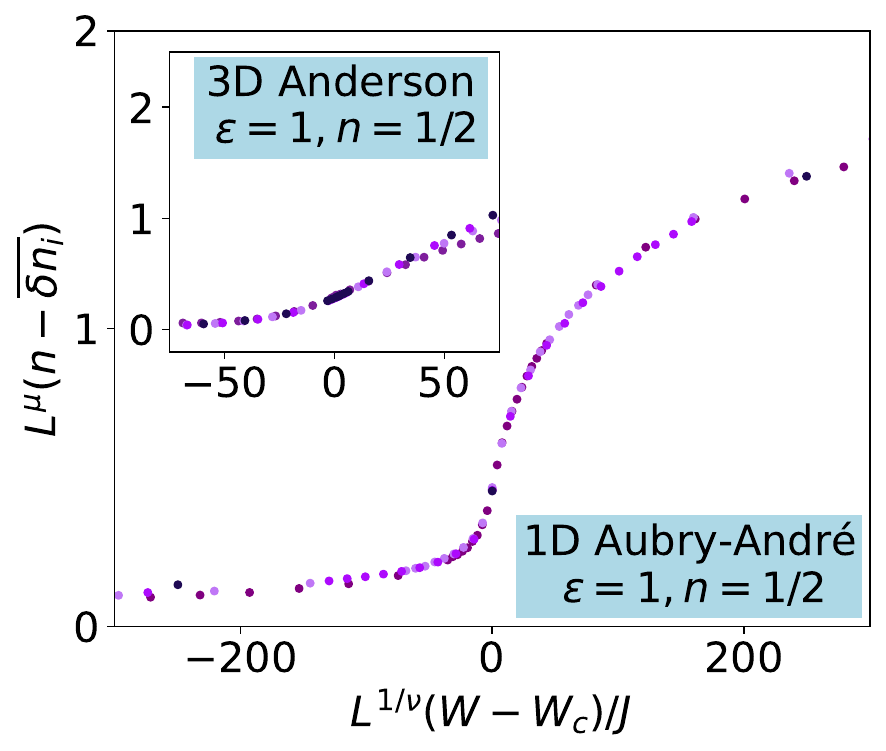}
\caption{{AAM: Scaling collapse of  $n-\overline{\delta n_i}$   
for  $L=256, 512, 1000, 20000$ and fitting parameters $W_c=2.00(5)J, \mu=0.25(5), \nu=1.05(5)$. 
Inset: Scaling collapse for the 3AM using 
$L^3= 12^3, 20^3, 28^3, 36^3$ with fitting parameters $W_c=16.3(2)J, \mu=0.50(5), \nu=1.60(5)$. 
}}
\label{difference_scaled1}
\end{center}
\vspace{-0.8cm}
\end{figure}

We next perform a scaling collapse of the data of Fig.$~$\ref{difference_raw_1DAA}, focusing on the 
delocalized phase and the vicinity of the transition point. For this purpose, in Fig.$~$\ref{difference_scaled1}, 
we replot  the data using the  dimensionless variables $w=L^{1/\nu}(W-W_c)/J$ and  $(n-\overline{\delta n_i})L^{\mu}$
 for the $x$ and $y$ axis, respectively. 
To get the best-quality estimate of the exponent $\nu$, the scaling collapse needs to be performed for system sizes in the scaling regime. Therefore, we include only systems with $L\geq256$, which is a posteriori justified by the results, see the inset of Fig.$~$\ref{difference_raw_1DAA}. For the AAM, we then find $\mu\simeq 0.25(5)$, $W_c/J=2.00(5)$ and the expected critical exponent $\nu\simeq 1.05(5)$, see the main panel of Fig.$~$\ref{difference_scaled1}. As a side remark, note that $\mu\simeq \alpha_c$, so that the product  $(n-\overline{\delta n_i})L^{\mu}  $ takes a finite value at  the transition point. Interestingly, we have recovered the values of $W_c/J=2$ and $\nu= 1$ known
from the single-particle transition \cite{Aubry80,Suslov82}, which suggests that the Fock-space delocalization-localization transition is driven by the divergent
localization (correlation) length in the single-particle states.

We carry out the same scaling-collapse procedure for the 3AM. Here, the best up-to-date estimate for the critical disorder strength is
 $W_c/J\approx 16.54$ \cite{Ohtsuki18}, which is obtained by the transfer matrix method for the mid-spectrum states of the single-particle Hamiltonian.
In the inset of Fig.$~$\ref{difference_scaled1}, we show the scaling collapse, obtained by retaining only the data sets
 with $L\geq12$, see the inset of Fig.$~$\ref{difference_raw_3DA}. 
 We then find $\mu\simeq 0.50(5)$, $W_c/J=16.3(2)$ and the expected critical exponent $\nu\simeq 1.60(5)$, see the inset of Fig.$~$\ref{difference_scaled1}. 
The estimates of the critical point $W_c/J\simeq16.3(2)$ and the critical exponent $\nu\simeq 1.60(5)$ are quite close to the values $W_c/J=16.54, \nu= 1.572$ obtained
by the transfer-matrix method \cite{Ohtsuki18}, where much larger system sizes (up to $L=64$) were considered. We attribute the small differences between the estimates to residual finite-size effects. Additionally, the many-body Slater determinant  can contain a fraction of single-particle states whose energy lies close to the band edge. Since these states localize for a much weaker disorder, it is in principle possible that $n-\overline{\delta n_i}$ saturates to a small 
but finite value in the vicinity of the critical point, in contrast to the scaling ansatz. For the system sizes that we considered, however, we do not find clear evidence for this effect.

The presented results for the AAM and the 3AM confirm that the occupation distance accurately captures the transitions well. 
Thus, remarkably, by exploiting Fock-space localization, the transition points can be determined from the distributions of the simple-to-calculate quantity $n_i$.
Moreover, the advances with characterizing MBL also feedback into devising hitherto unexplored approaches for disordered non-interacting models (see also \cite{Gullans2019,Suntajs2021,Prelovsek2021}).

\begin{figure}[t!]
\begin{center}
\includegraphics[width=8cm,angle=0]{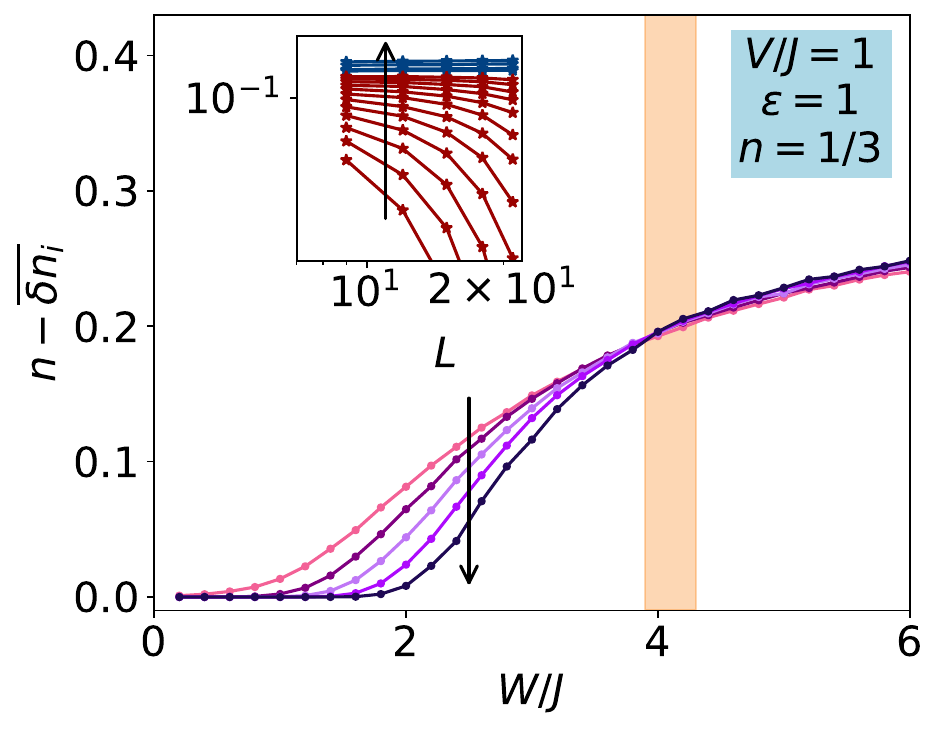}
\caption{1D interacting fermions, Eq.~(\ref{eq:ham}), ($\epsilon=1, V/J=1, n=1/3$): $n-\overline{\delta n_i}$ as a function of $W/J$ for $L=9, 12, 15, 18, 21$.
The   arrow   specifies   increasing  system size and the shaded area indicates the  estimated range of the transition point. Inset: 
 Log-log plot of the data from the main panel as a function of $L$   in the delocalized regime  (red symbols, $W/J= 1.2-3.6$, in steps of $0.2$)  
and in the localized phase  (blue symbols, $W/J= 4.2, 5.0, 6.0$). 
The   arrow   specifies   increasing disorder strength. } 
\label{interacting}
\end{center}
\vspace{-0.8cm}
\end{figure}

\section{Results for interacting models}

We now turn our attention to the interacting model in Eq.$~$\eqref{eq:ham}, considering values of the filling $ 1/10 \leq n \leq 2/3$,  $L \leq 30$, and $10^4$ 
disorder realizations (see the appendix for details). 
We impose periodic boundary conditions. For a given disorder realization, 
we use the shift-and-invert method \cite{Luitz14,Pietracaprina18} to efficiently
extract the six eigenstates closest to the target energy $\epsilon=1$.

The main panel of Fig.$~$\ref{interacting} shows $n-\overline{\delta n_i}$ 
 as a function of disorder strength for  $n=1/3$ (for other values of $n$, the behavior is similar).  
For $W/J<3.6$, there is a clear decay towards zero while for $W/J>4.2$, there is a slow saturation to a finite value. However, the decay in the delocalized phase
has a different character from the one previously observed for the non-interacting models. In particular, see the inset of Fig.$~$\ref{interacting}, the
decay of the  average occupation distance is not algebraic, but  exponential. This is a consequence of the exponential decay of the fluctuations
according to the eigenstate thermalisation hypothesis \cite{Deutsch91,Srednicki94,Rigol2008,Beugeling14}.

\begin{figure}[t!]
\begin{center}
\includegraphics[width=8cm,angle=0]{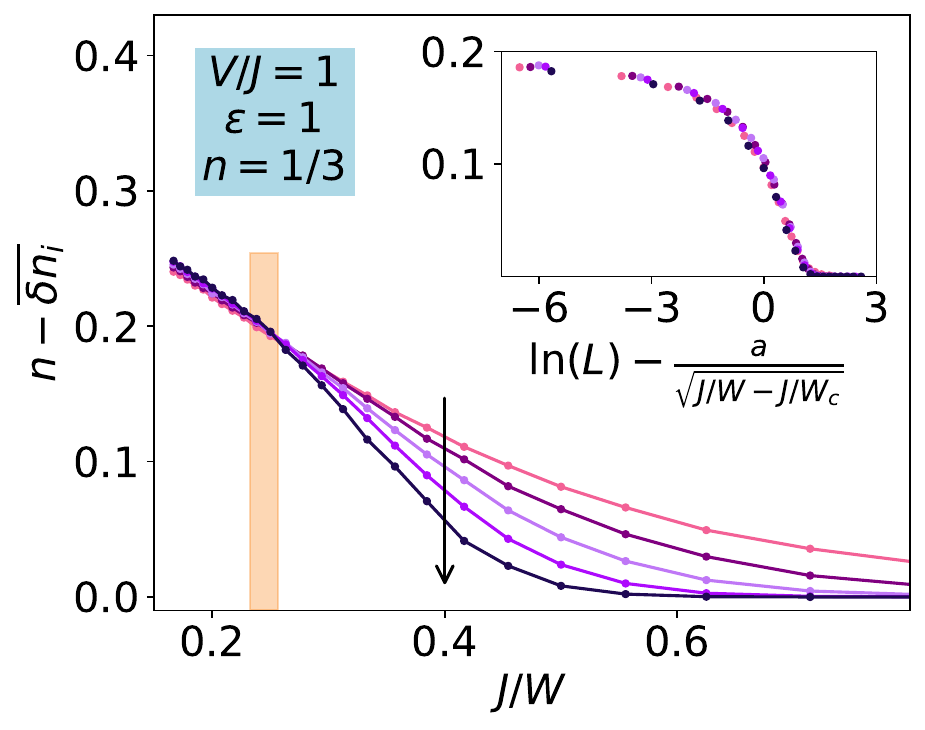}
\caption{1D interacting fermions, Eq.~(\ref{eq:ham}), with $\epsilon=1, V/J=1, n=1/3$: $n-\overline{\delta n_i}$  
as a function of $J/W$ for $L=9, 12, 15, 18, 21$.
The   arrow   specifies   increasing  system size and the shaded area indicates the estimated range of the transition point. Inset:
Scaling collapse of the data from the main panel in the delocalized phase as a function of 
$\ln(L/\xi)$ where $\xi=\exp(-a/\sqrt{J/W-J/W_c})$ with fitting parameters $W_c/J=4.1(2)$ and $a=1.00(5)$. For the scaling
collapse all the data in the ergodic regime, i.e. with $J/W>J/W_c$, are used.} 
\label{KT}
\end{center}
\vspace{-0.8cm}
\end{figure}

The exponential dependence on $L$ in the delocalized phase  suggests that, for the interacting system, 
the scaling collapse of the data in Fig.$~$\ref{interacting} is of the KT type. 
We also add that the KT form of the scaling is inspired by the recent literature on the phenomelogical
renormalisation group theories of the MBL transitions \cite{Goremykina19,Dumitrescu19,Morningstar19}.
To verify this hypothesis, in Fig.$~$\ref{KT} (main panel), we first display  $n-\overline{\delta n_i}$  as a function of  $J/W$. 
Then, in the inset of the same figure, we plot the {same} data as a function of the variable $\ln(L/\xi)$, where
$\xi=\exp(-a/\sqrt{J/W-J/W_c})$ is the KT length, with $W_c$ and $a$ being   fitting parameters.
Evidently, a  KT-like scaling  for the occupation distance is  {consistent}  with our data and the collapse is of similar quality in the
 whole delocalised region. We note that with the power-law scaling, we did not obtain any satisfactory collapse of our data (not shown).
Interestingly, a KT scaling predicts a jump of the occupation distance at the critical point $W_c$ for $L\rightarrow \infty$, which agrees with  the 
jump of the multi-fractal dimension of the eigenstates at the localization-delocalization transition reported in \cite{Mace20, DeTomasi20, Solorzano21,Roy21}.
Our analysis of {\em full} density distributions is advantageous over the study of the minimum deviation from $n_i=1$ -- a single number -- studied in \cite{Laflorencie20},
since our measure suffers less from numerical or experimental uncertainties.

\begin{figure}[t!]
\begin{center}
\includegraphics[width=7.5cm,angle=0]{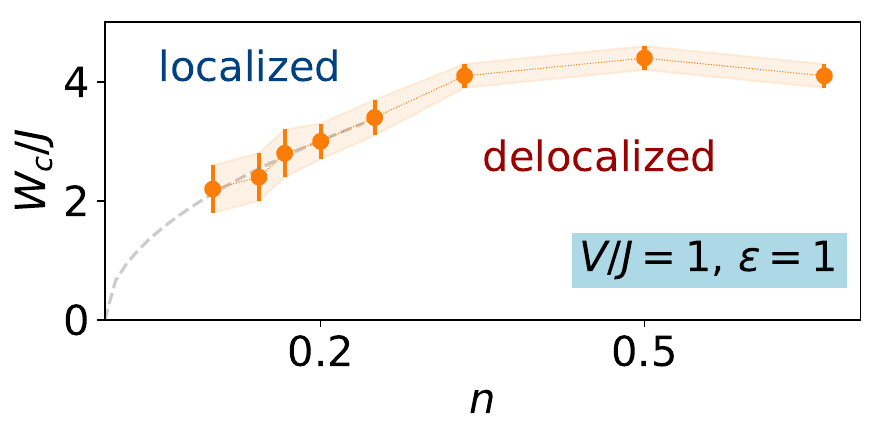}
\caption{1D interacting fermions  Eq.~(\ref{eq:ham}), with $\epsilon=1, V/J=1$:
Transition point $W_c/J$ as a function of  filling $n$ (dots) estimated from scaling collapses as 
in Fig.$~$\ref{KT}. The  dashed line is a power-law fit to $W_c \approx c\cdot n^{\lambda}$ 
where $\lambda = 0.58(10)$.} 
\label{estimation}
\end{center}
\vspace{-0.8cm}
\end{figure}

Finally, we repeat the same procedure for different particle fillings to extract $W_c=W_c(n)$. 
The full density dependence of the transition point in the interacting system was not explored before and
is an additional quantitative prediction which could be explored in future experiments. The density dependence of the transition in the 
interacting system is different from the AAM where the critical disorder strength is the same for all fillings, thus the $n$-dependence of $W_c/J$ hints at a different mechanism 
of the transition in interacting systems.
The resulting phase boundary between MBL and delocalized states is displayed in Fig.$~$\ref{estimation}.
First, note that the critical disorder strength is particle-hole symmetric with respect to filling, i.e., $W_c(n)=W_c(1-n)$. 
Second, $W_c$ takes its maximum value  at half-filling, where $W_c/J=4.4(2)$; 
consistent with other works \cite{Luitz14,Bera15,Mace20,Laflorencie20,Hopjan20} (see also \cite{Devakul15,Doggen18,Chanda20,Khemani17,Suntajs19,Sierant19a,Abanin19a,Panda20,Chanda20a,Suntajs20,Sierant20a}). 
Third, for $n<0.5$, $W_c/J$ decreases steadily as the  filling diminishes and must vanish at zero filling. Indeed, all single-particle states are 
localized in an infinite lattice and few-particle states remain also localized, although short-range interactions can substantially increase their localization length \cite{Shepelyansky:AndLocTIP1D:PRL94,vonOppen:AndLocTIPDeloc:PRL96,Frahm:EigStructAL1DTIP16} (a similar situation occurs also in two dimensions, see \cite{Stellin:PRB2020}).   
By fitting the data in Fig.$~$\ref{estimation} at low fillings  with a power law $W_c/J \approx c\cdot n^{\lambda}$, where $c$ and  $\lambda$ are fitting parameters, we obtain 
$\lambda = 0.58(10)$.

\section{Conclusions}
 We demonstrated that the quantitative analysis of density distributions is instrumental for the characterization  
of  localization-delocalization transitions. 
Our approach based on the occupation distance exploits both real-space {\it and} Fock-space localization as a characteristic properties of 
states with localized (quasi-)particles.
We showed that the average occupation distance \cite{Hopjan20}, extracted from the density distributions,
 exhibits  critical scaling behavior at  the transition of noninteracting models such as the 1D Aubry-Andr\' e  and the 3D Anderson model as well as  of interacting spinless fermions.  In the noninteracting models, the average occupation distance collapses with a power-law decay
 while in the interacting model, the observed KT scaling  hints at a different mechanism of the MBL
 transition consistent with predictions of \cite{Goremykina19,Dumitrescu19,Morningstar19,Suntajs20,Laflorencie20,Morningstar20}. 
Finally, we extract the filling dependence of the transition point and observe an approximate square-root dependence $W_c \propto \sqrt{n}$. 

The measurement of densities is simple and suffers less from errors in numerical methods such as DMRG than more complicated observables or time-dependent objects \cite{Schollwoeck11}.
Regarding quantum-gas experiments, while these do not access eigenstates, the measurement of distributions of densities after
slow loading processes into disorder potentials may yield similar information, to be studied further. 
Such experiments would neither  be  restricted to small systems  as is the case for the measurment of the entanglement entropy \cite{Lukin19,Chiaro19} 
nor
require time-dependent measurements, further reducing uncertainties.
Our results thus provide a path for studying delocalization-localization  transitions in future scaled-up quantum-gas microscope experiments,
aiming at clarifying the nature of the transition in interacting systems and the existence of the MBL phase.\\

\begin{acknowledgements}
We thank J. Bardarson and L. Vidmar for useful comments on an earlier version of the manuscript and for helpful discussions. 
\end{acknowledgements}

\appendix*

\section{Finite-size scaling ansatz and data collapses}

\begin{table}[t!]
\begin{tabular}{ |p{1cm}||p{0.9cm}|p{0.9cm}|p{1.1cm}|p{1.15cm}|  p{1.15cm}|  p{1.15cm}| }
 \hline
$n=\frac{2}{3}$ & L=9$~~$  (84)  &  12$~~$ (495)  & 15 (3003)   & 18 (18564) & 21 (116280) & \\
$n=\frac{1}{2}$   & L=10 (252)   & 12$~~$ (924) &   14 (3432) & 16 (12870) & 18 (48620) &\\
$n=\frac{1}{3}$ & L=9$~~$  (84)  &  12$~~$ (495)  & 15 (3003)   & 18 (18564) & 21 (116280) & \\
$n=\frac{1}{4}$ & L=12  (220)  &  16 (1820)  & 20 (15504)  && & \\
$n=\frac{1}{5}$ & L=15  (3003)  &  20 (4845)  & 25 (53130)&& & \\
$n=\frac{1}{6 }$ &  L=12  (66)  & 18$~~$ (816)  & 24 (10626)  && &\\
$n=\frac{1}{7}$ &  L=14 (91) & 21 (1330)  & 28 (20475)  && & \\
$n=\frac{1}{10}$ & L=20 (190)  &  30 (4060)  & & & & \\
 \hline
\end{tabular}
\caption{System sizes for the interacting system considered in this study, the corresponding sizes of the Hilbert space are written in the brackets. 
}
\label{table_of_sizes}
\end{table}

We now turn to the scaling collapses of $n-\overline{\delta n_i}$. For the non-interacting models, i.e., the AAM and the 3AM, we consider the scaling ansatz 
 of the  form $L^{-\mu} f(wL^{1/\nu})$ where $w=W/J-W_c/J$ is the distance from the transition point,
 $f{(t)}$ is the scaling function of interest, and $\{W_c, \mu, \nu\}$ are the fitting parameters. In the interacting model, we use an ansatz  
 of the KT form $g(\ln(L)-\ln(a/\sqrt{J/W-J/W_c}))$ with  {$g(t)$ being the} scaling function of interest and $\{W_c, a\}$ are the fitting parameters. 
 The system sizes considered for the interacting model are listed in Table$~$\ref{table_of_sizes}. We calculate 6 eigenstates close to $\epsilon=1/2$ for 
 $10^4$ disorder realization for Hilbert spaces $<25000$ and $10^2-10^3$ disorder realization for the larger ones. Note that this gives in total 
$L\cdot6\cdot10^4$ and $L\cdot6\cdot(10^2-10^3)$ values of the density $n_i$ for the density histograms which is sufficient for converged results for the averages.

To assess the quality of the scaling collapses, i.e., the smoothness of the functions $f(t)$ and $g(t)$ of the argument $t$, we use a cost function introduced in Ref.$~$\cite{Suntajs20}: 
\begin{equation}
\label{eq:measure}
\mathcal{C}_X= \frac{\sum_{j=1}^{N_p-1}|X_{j+1}-X_j|}{{\rm max}\{X_j\}-{\rm min}\{X_j\}} -1,
\end{equation}
where $X$ stands for the values of $f$ and $g$ and  the set $\{X_j\}$  of $N_p$ values $X_j$ is ordered by the values of the argument $t$.

\bibliographystyle{apsrev4-1}
\bibliography{paper_MBL_scaling}
\url

\end{document}